\documentclass[12pt]{article}
\usepackage{graphicx}
\newcommand{\ra}{\rightarrow}

\newcommand{\no}{\nonumber}
\newcommand{\be}{\begin{equation}}
\newcommand{\ee}{\end{equation}}
\newcommand{\ba}{\begin{eqnarray}}
\newcommand{\ea}{\end{eqnarray}}

\newcommand{\dta}{\mbox{$\delta$}}

\newtheorem{rmk}{Remark}[section]
\begin{document}
\thispagestyle{empty}
\title{ 
Modeling Vocal Fold Motion with a New Hydrodynamic Semi-Continuum
Model}
\author{M. Drew LaMar, 
\thanks{Department of Mathematics, University of Texas at Austin,
Austin, TX 78712.}
Yingyong Qi 
\thanks{ Qualcomm Inc, 5775 Morehouse Drive, San Diego, CA 92121.}
and Jack Xin 
\thanks{Department of Mathematics, University of Texas at Austin,
Austin, TX 78712. Corresponding author, email: jxin@math.utexas.edu.
This work was partially supported by ARO grant DAAD 19-00-1-0524.
 }}
\date{}
\baselineskip=18pt
\maketitle

\begin{abstract}
Vocal fold (VF) motion is a fundamental process in voice production, 
and is also a challenging problem for direct numerical computation 
because the VF dynamics depend on nonlinear coupling of air flow with the 
response of elastic channels (VF),
 which undergo opening and closing, and induce
internal flow separation. A traditional modeling approach 
makes use of steady flow approximation or Bernoulli's law which is 
known to be invalid during VF opening.   
We present a new hydrodynamic semi-continuum system for VF 
motion. The airflow is modeled by a quasi-one dimensional 
continuum aerodynamic system, and the VF by 
a classical lumped two mass system. The reduced flow system 
contains the Bernoulli's law as a special case, and is 
derivable from the two dimensional 
compressible Navier-Stokes equations. Since we do not make steady flow  
approximation, 
we are able to capture transients and rapid changes of solutions, 
e.g. the double pressure 
peaks at opening and closing stages of VF motion consistent with  
experimental data. We demonstrate numerically that 
our system is robust, and models in-vivo VF oscillation 
more physically. It is also  
much simpler than a full two-dimensional Navier-Stokes system.   
\end{abstract}
\vspace{.6 in}

\hspace{.2 in} PACS numbers: 43.70Bk, 43.28Ra, 43.28Py, 43.40Ga.
\section{INTRODUCTION}
Vocal folds (VF) are the source of the human voice, and their motion 
is a fundamental process in speech production. 
In recent years, mathematical modeling of vocal folds has been 
pursued as a viable
alternative to direct experimental studies using 
strobovideolaryngoscopy or electroglottography techniques.
Numerical simulations of VF models then provide us with a valuable
tool to understand, monitor and predict various behaviors of
normal and disordered voices in vivo. Together with models of vocal tract,
one can construct a voice simulator which clinicians, speech therapists, 
voice teachers,
and otolaryngologists can use to help 
with their skill improvement, diagnosis and patient treatment.
\bigskip

Since VF motion is mechanical and results from the
nonlinear interaction of airflow and elastic response of VF,
partial differential equations (PDEs)
can be written down from classical continuum mechanics
based on our knowledge of VF structures  
and air flow characteristics.
However, the complexity involved is daunting, 
both in terms of airflow and VF structure, 
for a direct simulation of a complete set of  governing equations. Also it is not necessary that one needs all the details of
such a solution to describe the main VF properties.  
Modeling effort is required
 to build a smaller set of equations that can capture the essential features of
VF dynamics.
In the past decade, much progress has been made in
modeling the elastic aspect of VF.
There are by now a hierarchy of
elastic models for VF, from the two mass model of
Ishizaka and Flanagan \cite{Ishi}, Bogaert \cite{Bo}, to 16 mass as well
as the continuum model of Titze and coworkers \cite{ST,T73,T74,Alip0}. 
However, the modeling of airflow or the fluid
aspect of VF is much less explored. 
\bigskip

There are broadly two types of approaches in treating the glottal flow.
One approach is to combine the Bernoulli's law 
in the bulk of the flow (steady flow approximation) with empirical formulas in
boundary layer, flow separation and wake 
\cite{Ishi,Bo,ST,Pel1}.
This approach oversimplifies the flow in the sense that
{\bf PDEs are approximated by algebraic equations}.
Though the approach is a working method
for building simple models, it clearly introduces
drastic approximations.
For example, it was realized \cite{Flan1} and concluded \cite{Mong}
that Bernoulli's law is not
valid during one-fifth of the VF vibration cycle, especially at
VF opening and closure. This
lack of accuracy as a result of deviating significantly from the original PDEs
 is a major drawback of the empirical approach.
The other approach is to directly simulate
the two dimensional
Navier-Stokes (NS) system. Two dimensionality is a common 
assumption for vocal flows 
\cite{Pel1}. Alipour et al. formulated a steady state simulation with
a given glottal geometry \cite{Alip1}. Both \cite{Pel1,Alip1} appeared to have
been done for fixed channel shape, or in other words, in-vitro VF. 
To accurately model the pulsating nature of the flow during VF vibration 
however, a time dependent solution is more appropriate.
Yet in such a case, it is highly difficult to
resolve the flows in the presence of moving boundaries, closures,
and flow separation. Existing works are few in
this direction although a lot of measurements on the
flow characteristics such as intraglottal air pressure
and flow velocity have been made,  by
Titze and others \cite{Pel1,Mong,Moore,Ti2,Jiang,Aus,Hsi,Alip1,Stevens}.

\bigskip

The current status of flow modeling calls for 
a systematic study of
reduced PDE flow models and their coupling
with existing elastic models. In this paper, we develop 
an intermediate in-vivo PDE model system
so that {\bf original airflow PDEs are approximated by
reduced PDEs, not
algebraic equations}. Consider for now that the two sides of VF
are symmetric to each other across the centerline, 
the methodology could be extended 
to the asymmetric case.
The air flow is modeled by a quasi-one 
dimensional (vertical or upward direction) system of flow equations. 
The flow variables (pressure, velocity,
 density) are averaged quantities over the channel 
cross section of the corresponding ones in two dimensional NS system.
Assuming that the flow is predominantly in the 
vertical direction, which is reasonable 
before flows become turbulent in the exit 
region,  we derive the model flow system 
from the two dimensional isentropic 
compressible Navier-Stokes equations, see the appendix, also \cite{XHQ}.
If the channel is not changing in time, the system 
reduces to the familiar quasi-one 
dimensional gas dynamic equations in studying 
duct flows in aerodynamics
\cite{Liu1,Liu2,Whith,Men}. If the 
channel varies in time, there is an 
additional source term in the flow momentum 
equation, which turns out to be essential for 
drawing connection with the Titze
theory of small VF oscillation \cite{Ti0}.
One can regard the reduced airflow 
system as a coarse-grained NS system
which contains the Bernoulli's law as a special case, 
and inherits time dependent convection 
mechanisms from the full two dimensional 
NS system. The advantage is that the system is able to handle
time dependent effects, such as rapid pressure and 
velocity changes, during VF opening and closing; moreover, it is a lot simpler to simulate numerically
because all the unknown dependent 
variables are one dimensional in space.
Such a system will be coupled to an improved two mass model
\cite{Bo} for the VF cross section area 
to form a complete VF model. The VF cross section 
area appear as variable 
coefficients in the quasi-one dimensional air flow system. 
The VF motion is described
by how VF cross section area varies in time.

\bigskip 

The rest of the paper is organized as follows. In section 2, we introduce 
the equations of the model, and address related modeling issues. 
In section 3, we discuss numerical method and numerical results of 
model simulation. We show numerically that our model is able to 
generate VF motion in-vivo, and 
recover several known VF characteristics supported by experimental 
measurements, for example, unequal double pressure peaks at VF opening and
closure. We also show the robustness of our model by varying 
subglottal pressure and plotting how air volume velocity (air flux) 
changes as a function of time. Our results reach complete qualitative 
agreement with existing 
VF flow data. The conclusion is in section 4, and 
acknowledgement in section 5. Section 6 is the appendix 
on the derivation of our reduced flow model from the two dimensional 
compressible Navier-Stokes system. Table 1, figure captions and 
figures follow the references.  

\section{THE SEMI-CONTINUUM MODEL}
\setcounter{equation}{0}
Suppose the larynx is a two dimensional channel with a 
finite mass elastic wall of cross section width
$A(x,t)$, and length $2L$. The vocal fold is lumped into a sum of 
two masses connected by a spring, and each mass is connected to 
solid wall by a spring and a damper, the classical scenario 
in the two mass model \cite{Ishi,Bo}.
The air flows from $x= -L$ to $x
=L$, and is modeled by 
the quasi-one dimensional system derived in \cite{XHQ}: 

\noindent $\bullet$ conservation of mass: 
\begin{equation}
( A\rho )_{t} + (\rho\, u \, A)_{x} =0,  \label{E1}
\end{equation}
$\rho$ air density, $u$ air velocity;

\noindent $\bullet$ reduced momentum equation: 
\begin{equation}
(\rho uA)_{t} + (\rho u^2 A)_{x} = - (pA)_{x}+ A_x p + {\rho u A_t},
\label{E2}
\end{equation}
$p$ air pressure.

Assuming that the temperature is maintained 
as constant, so the airflow is isothermal, then the equation of state is:
\be
p= a^2 \rho, \label{E3}
\ee
where $a$ is the speed of sound. The cross section width $A$ is 
a piecewise linear function in $x$ determined by the locations 
of the two masses ($y_1$, $y_2$), in the classical two-mass 
model system (Bogaert \cite{Bo}, Ishizaka and Flanagan \cite{Ishi}):
\ba
m_1 y_{1}'' &+& r_1 y_{1}' +k_{1}(y_1 -y_{0,1}) + k_{12}(y_1 -y_2 +y_{0,12})
=F_1, \label{E4}\\
m_2 y_{2}'' & +& r_2 y_{2}' +k_{2}(y_2 -y_{0,2})+k_{12}(y_2-y_1-y_{0,12})
=0, \label{E5}
\ea
where $F_1= L_g \int_{-L}^{x_s}\, p\, dx$, $L_g$ the transverse (to the flow) 
dimension of vocal fold;
$y_i$'s are VF openings at locations $x_i$'s, $-L < x_1 < L= x_2$;
$x_s=x_2$ if there is
no flow separation, and $x_s =$ the location of flow separation if
it occurs.
The $m_i$, $r_i$, $k_i$, $i=1,2$, are 
mass density, damping and elastic spring constants. 
Mass one (lower mass) is situated near the VF entrance, and mass 
two (upper mass) is located 
towards the exit of the glottal region.
Following Bogaert\cite{Bo}, $x_s$ will be estimated by an empirical formula 
on the degree of divergence of the VF. Our complete VF model 
is the coupled system (\ref{E1})-(\ref{E5}).  

To make the paper self-contained, the derivation of (\ref{E1})-(\ref{E2}) 
is included in the appendix.
The flow variables in the quasi-one dimensional 
system are averages over the channel 
cross section of the corresponding ones in 
two-dimensional flows. The viscous effect in the flow produces 
the term $\rho u A_t$ from the no-slip boundary condition of the 
two-dimensional flows. Without this term, the above system 
is a familiar one used in gas dynamics (see \cite{Whith,Liu1,Liu2,Men} 
and references therein) for modeling flows 
through ducts with variable cross section. It is shown \cite{XHQ}
 that {\it this 
extra term is critical in transferring energy from airflow into the VF}, 
as the Titze theory \cite{Ti0} predicted. We have ignored the 
viscous terms in the momentum equation for simplicity, 
they appear to be higher order. 

Two mass model 
(\ref{E4})-(\ref{E5}) is a recent improvement \cite{Bo} of the original 
IF72 \cite{Ishi} in that flow separation point is not always 
at the VF exit, instead it depends on the glottal geometry. 
 Flow separation 
basically refers to a change of flow behavior from being attached to 
the VF cover via a viscous boundary layer to a developed free jet 
with vortical structures and turbulent wake. Because of the 
vortical buildup, pressure near the wall is typically low, and 
can be approximated by setting it to zero (or ambient 
pressure) as done on mass two in (\ref{E5}) when there is no vocal 
tract.
In converging glottis, there is no flow separation, however in 
diverging glottis, it occurs if the diverging angle is large enough. 
It is as yet a challenging problem (no simple theoretical prediction) to
decide for a flow when and where separation occurs. 
It is expedient for modeling purpose to adopt a working 
hypothesis supported by experiments \cite{Bo,Pel1}: 
\ba
& & y_2/y_1 < 1.1 \Longrightarrow \;\; x_s=x_2, \label{E6}\\
& & y_2/y_1 > 1.1 \Longrightarrow \;\; x_s=x_1 +
 {(x_2 -x_1)y_1 \over 10(y_2 -y_1)}, \; y_s = 1.1 y_1. \label{E7}
\ea
Notice that the flow separation location is a variable depending on 
the diverging angle.
It is worth pointing out that the assumptions we made for deriving 
the reduced flow model are all valid prior to the separation point. 
We expect to see a deviation after the flow separation point between 
the reduced flow model and the fully two dimensional NS solutions; 
however, flow pressure post separation is not used in (\ref{E5}). 
Thus our reduced flow model matches perfectly with the 
improved two-mass model \cite{Bo}. 

We also adopt the elastic 
collision (stopping) criterion in \cite{Ishi,Bo} 
when the two sides of VF approach each other 
and close. When $y_i$'s are smaller than a critical level 
$y_c$, then VF is considered closed. Following \cite{Ishi,Bo}, 
$(m_i,r_i,k_i)$ ($i=1,2$) are adjusted to closure 
values. In this case, the flow equations are solved only over 
$x \in [-L,x_1]$, and in (\ref{E4})-(\ref{E5}) the pressure force 
is adjusted to $F_1 = L_g \int_{-L}^{x_1} p\, dx$. Due to constant input 
pressure $p_0$, pressure at $x_1$ builds up. 
The two mass ODE's (ordinary differential equations) 
are still running even during VF closure, and 
in due time the increased pressure reopens VF.

The VF model system is posed as an initial boundary value problem on
$x \in [-L,L]$, with inlet boundary condition 
$(\rho,u)(-L,t)=(\rho_0,u_0)$, and a zero Neumann type boundary 
condition at exit $(\rho_x,u_x)(L,t)=0$. 
The advantage of such Neumann  
type boundary conditions is 
that it helps the flow to go out of the computational domain, which 
is needed for a stable numerical method
free of numerical boundary artifacts. Our numerical experiments suggest
that the above treatment works fine.

The major difference between our model and that of Bogaert \cite{Bo} is 
that we do not make quasi-steady approximation on the flow variables, 
instead we integrate time dependent system (\ref{E1})-(\ref{E2}). 
This turns out to be particularly important for capturing 
transients near closure and reopening stages of VF motion.

It is helpful to put the system (\ref{E1})-(\ref{E2}) into 
a rescaled form. Let $v=u/a$, $a$ the speed of sound. Then:
\ba
{1\over a}(Ap)_t & + & (p v A)_x =0, \no \\
{1\over a}(pvA)_t & +&  (pv^2 A)_x = -(pA)_x +A_x p +pv A_t /a, \label{E8}
\ea
where typically $v=u/a \approx 0.1$, the Mach number. If we use the 
convenient cm-g-ms unit, $a= 35 \, cm/ms$, $1/a $ is a small parameter. 
If we ignore the terms with $a$, we have exactly Bernoulli's law 
for steady flows. However, these seemingly small terms are 
essential especially during opening stage of VF, and should be kept 
for an accurate time-dependent solution.   

%
%
%
%
%
%
%
%
%
%
%
%
%
%
%
%
%

\section{NUMERICAL METHOD AND SIMULATION RESULTS}
\setcounter{equation}{0}
For given VF shape, $A(x,t)$,  the flow system (\ref{E1})-(\ref{E2}) 
is of the form:
\be
U_t +(F(U))_x = G(U), \label{E9}
\ee
so called conservation law (see \cite{Lev} and references therein)
with lower order source term $G$. The
function $F$ is the flux function. 
We implemented a first order finite difference method, where time marching
is split into two steps. In the first step ($t=nk \ra (n+{1\over 2})k$),
we solve the conservation law
$U_t+(F(U))_x=0$ with explicit Lax-Friedrichs method \cite{Lev}:
\be
U_{j}^{n+{1\over 2}} = {1 \over 2}(U_{j-1}^{n} + U_{j+1}^{n}) - {k \over 2h}
\left(F(U_{j+1}^{n}) - F(U_{j-1}^{n})\right), \label{E10}
\ee
where $k$ and $h$ are time step and spatial grid size. Here $k$ must be
small enough to ensure stability of the difference scheme and
to keep the computed flow
velocity positive (no back flow is allowed). In step two
($t+{k\over 2}\ra t +k$),
we update the solution from $U^{n+{1\over 2}}$ to $U^{n+1}$
by implicitly integrating ODEs: $U_t =G(U)$ in
flow equations, and the two mass equations (\ref{E4})-(\ref{E5}); where we
apply central differencing in space and backward differencing in time.
In the first step, $U$ is updated using VF shape $A$ at time $t=nk$;
in the second step, the ODEs from two mass system and
source terms are solved to update
solutions to $(n+1)k$. We point out that when VF approach closure,
the ODE's in step two become rather stiff, and this is the main
reason to use implicit backward differencing in time \cite{GO}.
The time step $k$ is a variable. It is smaller when VFs are closing and 
the equations are stiff,  
and is larger when VFs are opening up. 
\bigskip

The parameters used in our calculation are: space grid size $h=0.01125$,
variable time step $k \in (10^{-6},10^{-4})$. 
The time unit is ms, length unit cm, 
$2L=0.225 \, cm$, speed of sound
$a =35 \, cm/ms$, $u_0=4 \, cm/ms$, $p_0=7840$ $dynes$ $ cm^{-2}$. 
Other two mass model parameters are in Table 1. In runs not shown, we 
have reduced $h$ to half or even smaller sizes, and observed 
similar findings as reported below.

\bigskip

\begin{figure}[ht]
\centerline{\includegraphics[width=350pt, height=340pt]{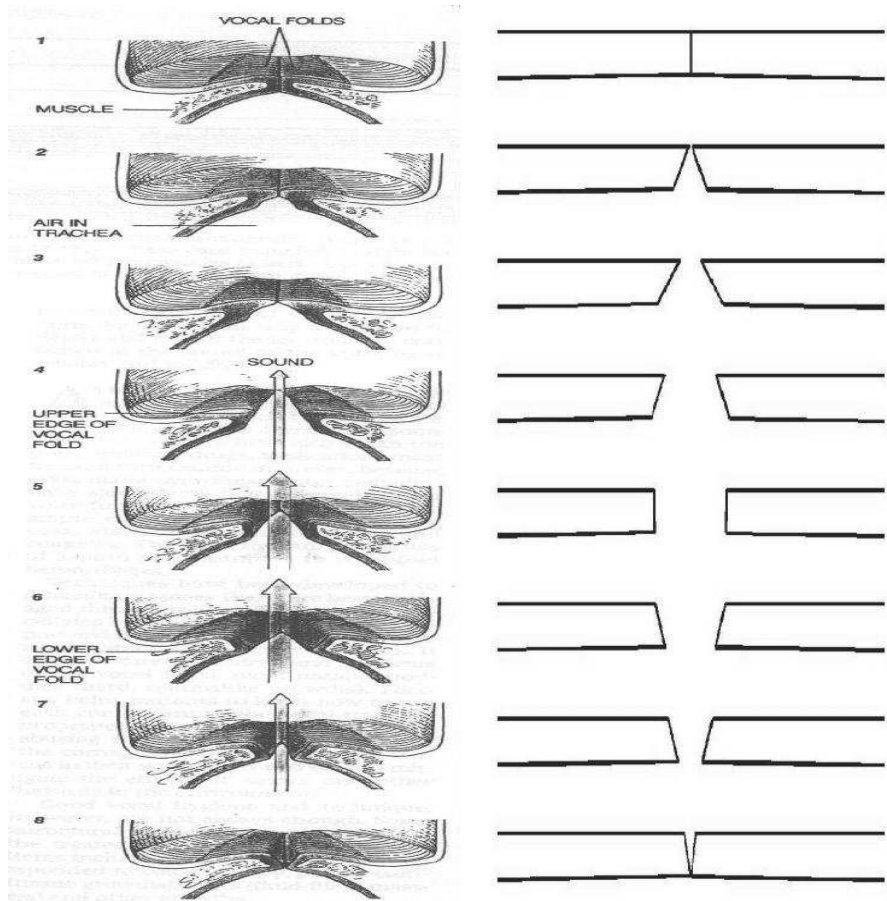}}
\caption{VF Vibration: left column --- p. 113 of Sataloff \cite{Sat},
right column --- simulated VF with our model (\ref{E1})-(\ref{E5}).
}
\label{Fig1}
\end{figure}

Now we describe our numerical results, and compare with figures in
the literature or those from experimental measurements.
In Figure \ref{Fig1}, we show a comparison of a cycle of
VF vibration. The left column is the figure on page 113 of
Sataloff's Scientific American article \cite{Sat}, the right column
is a plot of our numerically simulated VF vibration cycle. The
resemblance is clear. A web animation is also available \cite{LQX}.

\bigskip

\begin{figure}
\centerline{\includegraphics[width=350pt, height=340pt]{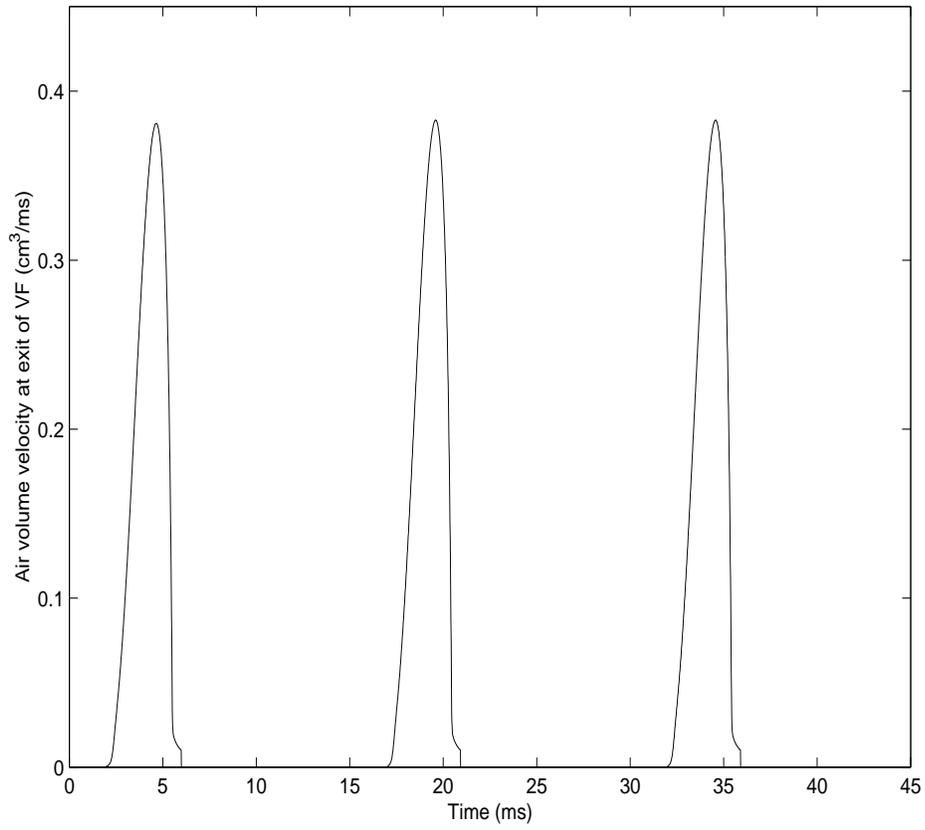}}
\caption{Simulated VF volume velocity (air flux, $cm^3/ms$ ) vs time at exit
of VF from model (\ref{E1})-(\ref{E5}).}
\label{Fig2}
\end{figure}

In Figure \ref{Fig2}, we show our computed air volume velocity 
(air flux) at the exit of VF, which
compares well with Fig 6, Fig 7, Fig 8 of Story and Titze \cite{ST}.
Notice that the airflux goes down to zero gradually at VF closing 
then drops down abruptly to zero. The drop is due to the cutoff 
introduced in the two mass model for closure, the $y_c$. Below 
$y_c$, the cross section area $A$ becomes very small, and  
the flow calculation becomes rather stiff. In other runs  
(not shown), 
we have observed that increasing $y_c$ will 
shorten the curved portion and straighten up the plot near closures 
($t\approx 6, 21, 36$ ms). The air volume velocity shows asymmetry, 
steeper on the right side than on the left side, consistent with  
Fig. 3b of Titze \cite{Ti0}. 

\bigskip

\begin{figure}
\centerline{\includegraphics[width=350pt, height=340pt]{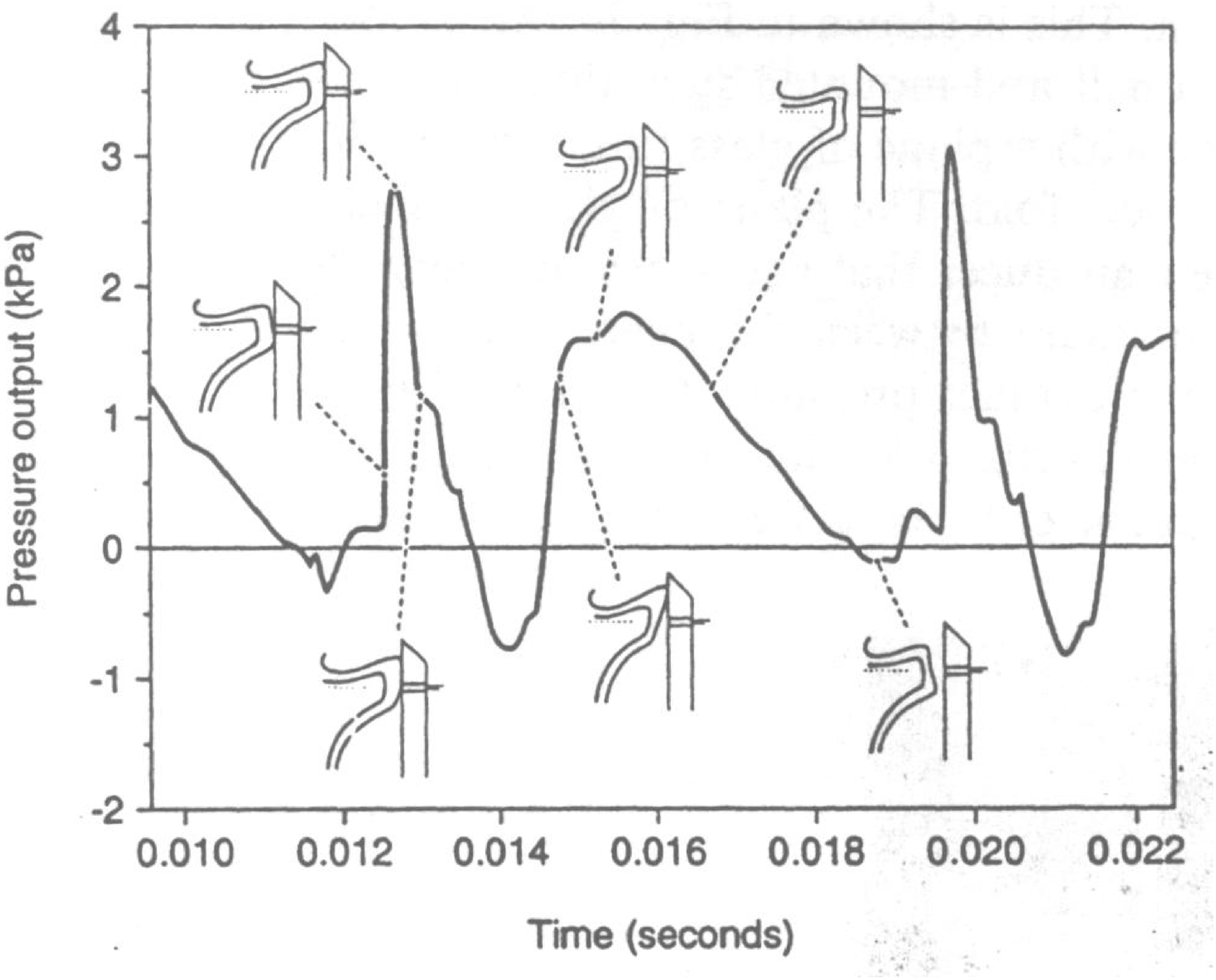}}
\caption{Experimentally measured intraglottal pressure on excised canine
larynx, see Fig. 8 on page 426 of Titze \cite{Ti2}.
}
\label{Fig4}
\end{figure}

\begin{figure}
\centerline{\includegraphics[width=350pt, height=340pt]{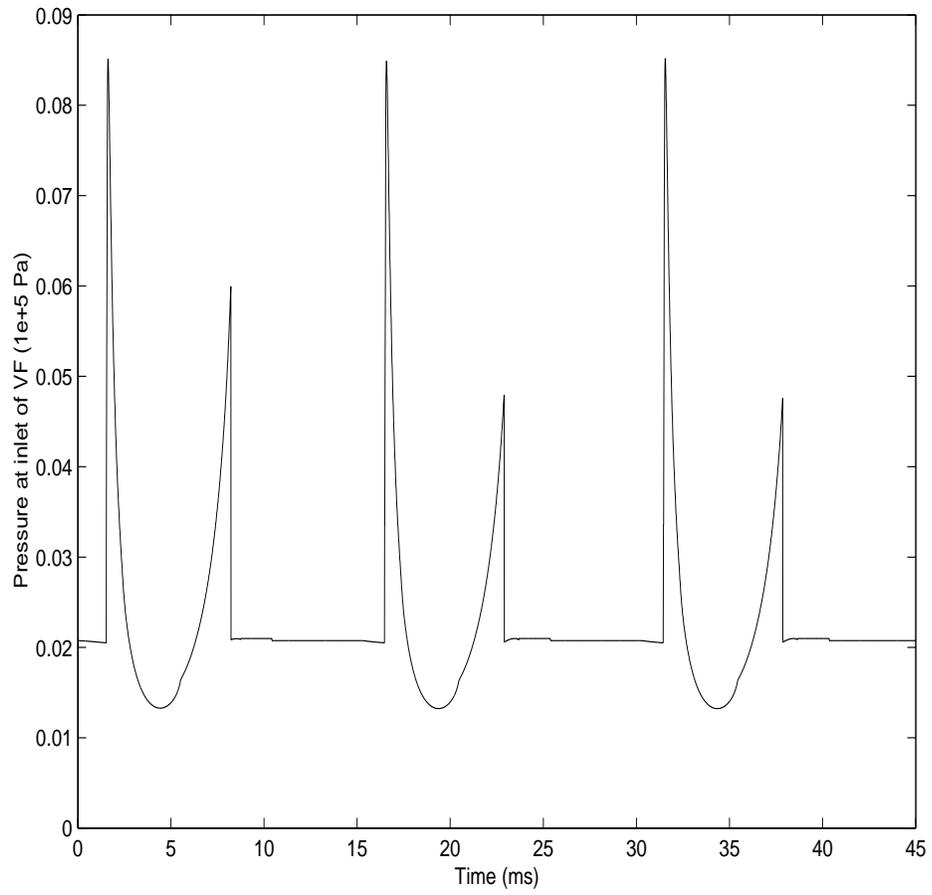}}
\caption{Computed pressure ($10^5$ Pa) change in time at
the point right before lower mass.
}
\label{Fig5}
\end{figure}

Figure \ref{Fig4} is the experimentally measured intraglottal pressure
on an excised canine larynx from \cite{Ti2} (Fig. 8 there) and \cite{Jiang},
which showed the double peak (intraglottal) pressure structure
respectively at VF opening and closing. Figure \ref{Fig5} is our computed
pressure at the grid point before lower mass.
The double peaks are present and
resemble well those in Figure \ref{Fig4}, only that ours are
steeper and higher. Several factors contribute to the difference:
 (1) we used inviscid flow model while there was physical air viscosity
in experiments that smooth the solutions, (2) the closure treatment
of two mass model differs from the actual
VF closure, (3) Figure \ref{Fig4} plots the pointwise pressure, not 
an average pressure over glottis. 
At the qualitative level however, our model solutions are
in full agreement with the experimental finding.
Notice that the second peak is lower than the first.

\bigskip

To the best of our knowledge, the experimentally observed
unequal double pressure peaks \cite{Ti2,Jiang}, have not been computed
previously in a VF model without coupling to vocal tract.
The experiment had no vocal tract load.
In Story and Titze \cite{ST}, 
a computed two peak intraglottal pressure plot was given
(see Fig 11 \cite{ST}) using their three mass VF model; however,
there is additional coupling to
vocal tract or an additional subglottal system. Also their
computed second peak appeared higher than the first peak.
{\it The fact that our model can recover the experimental double
pressure peaks renders strong support for its validity and value.}

\bigskip

\begin{figure}
\centerline{\includegraphics[width=350pt, height=340pt]{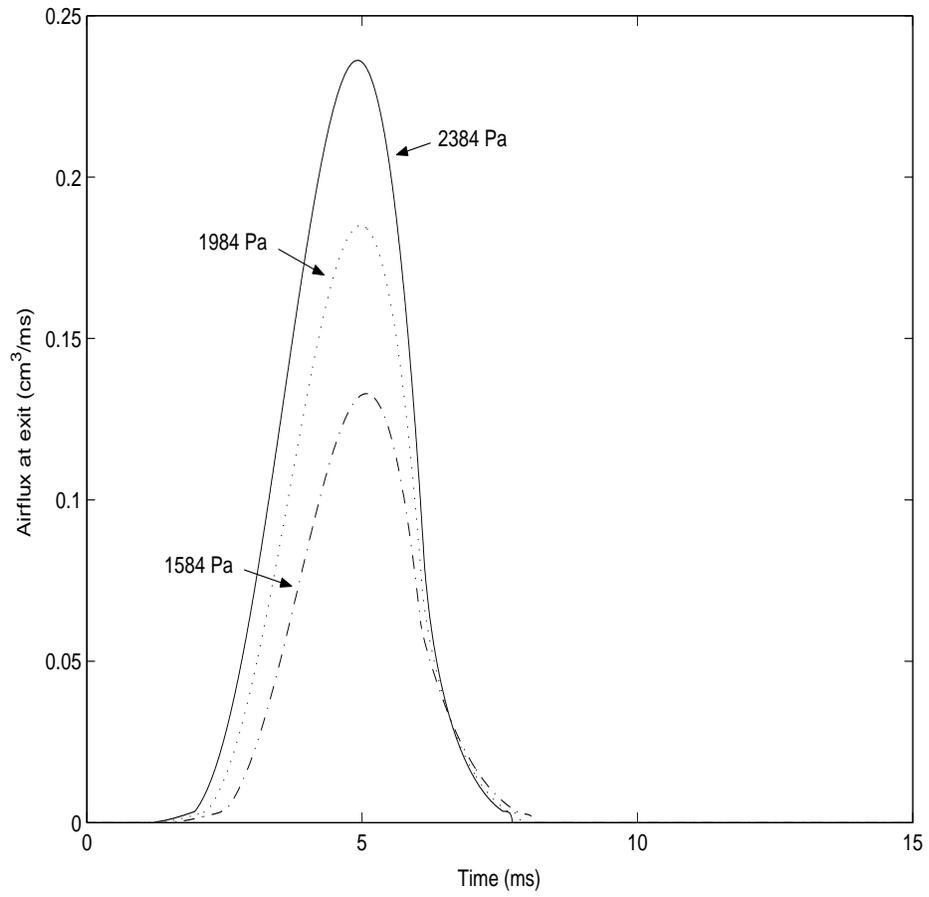}}
\caption{The computed air volume velocity for three values of
subglottal pressures.
}
\label{Fig5a}
\end{figure}

We also tested our model robustness under pressure
variation.
In Figure \ref{Fig5a}, we show a plot of air volume velocity vs time at
VF exit for three subglottal pressures: 1584 Pa,
1984 Pa, 2384 Pa with other parameters same as in Table 1.  
We see that as subglottal pressures increase
with other parameters fixed, air volume velocity
curves get higher (at peaks) and steeper (at two sides).
{\it This agrees very well with Fig. 2.14(a), page 78,
 of K. Stevens \cite{Stevens}, and is another strong support
for our model.}

\bigskip

\begin{figure}
\centerline{\includegraphics[width=350pt, height=340pt]{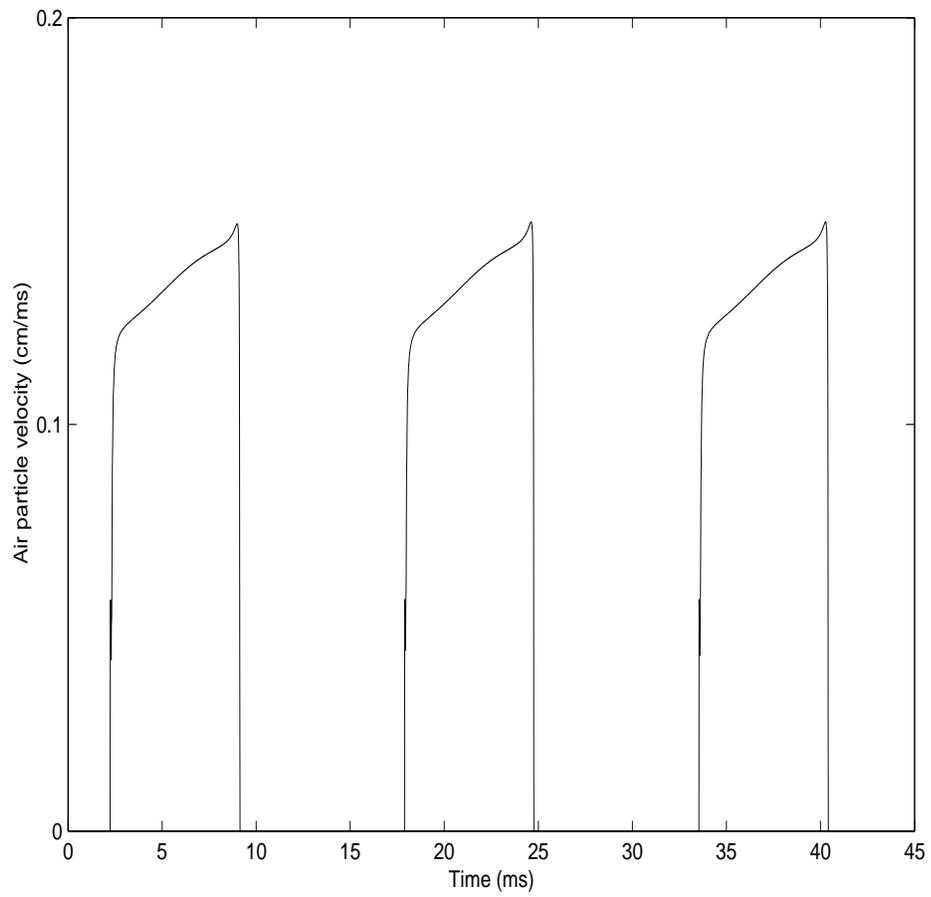}}
\caption{The computed air particle velocity vs. time.
}
\label{Fig5b}
\end{figure}

In Figure \ref{Fig5b}, we show an air particle velocity at VF exit
(after upper mass) over three vibration
cycles for subglottal pressure 2384 Pa, which agrees qualitatively
with Fig 3c, page 1538, of Titze \cite{Ti0}. For Figure \ref{Fig5b}, 
a small amount of additional numerical diffusion is added in 
(\ref{E1})-(\ref{E2}).

\section{CONCLUDING REMARKS}
In this paper, we introduced a new semi-continuum VF model consisting 
of a reduced PDE flow system \cite{XHQ} 
and a recent two mass elastic system \cite{Bo}.  
The flow part of the model is more physical than a traditional 
treatment with Bernoulli's law yet much simpler than a full 
two dimensional Navier-Stokes system. The reduced PDEs are 
derivable from the two dimensional compressible 
Navier-Stokes system, and are much more economical for computation.
We demonstrated numerically that the model solutions are in 
qualitative agreement with known VF experimental measurements. 
In future work, we plan to couple the flow model with more 
physiological elastic VF models, such as \cite{ST}; compute with 
higher order finite difference methods \cite{Lev} for attaining 
more accuracy;  
analyze qualitative properties (phonation thresholds) of model solutions 
using bifurcation methods; and incorporate additional viscous effects in the 
flow. 

\section{ACKNOWLEDGEMENTS}
The authors wish to thank Profs. I. Titze and F. Alipour for 
helpful conversations and email communication on VF modeling, and for 
their recent work \cite{Alip0}. The authors would like to thank 
Dr. J. M. Hyman for suggesting references \cite{Men,Sat}. 

\section{APPENDIX: DERIVATION OF REDUCED FLOW SYSTEM}

\setcounter{equation}{0}

We derive the fluid part of the model system assuming that the fold varies
in space and time as $A=A(x,t)$. Consider a two dimensional slightly viscous
subsonic air flow in a channel with spatially temporally varying cross
section in two space dimensions, $\Omega_0 =\Omega_0(t) = \{ (x,y): x \in
[-L,L], y \in [-A(x,t)/2,A(x,t)/2]\}$, where $A(x,t)$ denotes the channel
width with a slight abuse of notation, or cross sectional area since
 the third dimension is uniform. The two dimensional Navier-Stokes
equations in differential form are (Batchelor \cite{Batch}, page 147):

\noindent $\bullet$ conservation of mass: 
\begin{equation}
\rho_{t} + \nabla \cdot (\rho\, \vec{u}) =0;  \label{A5}
\end{equation}

\noindent $\bullet$ conservation of momentum: 
\begin{equation}
(\rho \vec{u})_t = - \nabla \cdot (\rho\, (\vec{u}\otimes \vec{u})) +
div ( \sigma \cdot \vec{n}\, );  \label{A6}
\end{equation}
where $\sigma$ is the stress tensor, $\sigma =(\sigma_{ij})= - p %
\mbox{$\delta$}_{ij} + d_{ij}$, and: 
\[
d_{ij}= 2\mu\, (e_{ij} - {\frac{div \vec{u} }{3}}\mbox{$\delta$}_{ij}),\;\;
e_{ij} = {\frac{1}{2}}(u_{i,x_{j}} + u_{j,x_{i}}), \;\; (x_1,x_2) \equiv
(x,y); 
\]
$\mu$ is the fluid viscosity; $\Omega (t)$ is any volume 
element of the form: ($\vec{u}=(u_1,u_2)$)
\be
\Omega (t)= \{ (x,y): x \in [a,b] \subset [-L,L], y \in [-A(x,t)/2,A(x,t)/2].
\}. \label{dom} 
\ee
The equation of state is either polytropic or isothermal.

The boundary conditions on $(\rho,\vec{u})$ are:

\noindent (1) on the upper and lower boundaries $y = \pm A(x,t)/2$, $%
\rho_{y} = 0$, and $\vec{u}=(0,\pm A_{t}/2)$, the velocity no slip boundary
condition;

\noindent (2) at the inlet, $x=-L$, $p=p_0$, given subglottal pressure, 
$(u_{1},u_2)=(u_{1,0},u_{2,0})$, given input flow velocity. At the exit. 
$(p,u_1,u_2)_{x}=0$, to help the waves go out of the 
domain freely.

We are only concerned with flows that are symmetric in the vertical. For
positive but small viscosity, the flows are laminar in the interior of $
\Omega_0$ and form viscous boundary layers near the upper and lower edges.
The vertically averaged flow quantities are expected to be much less
influenced by the boundary layer behavior as long as $A(x,t)$ is much larger 
than $O(\mu^{1/2})$. We also ignore effects of possible flow seperation
inside $\Omega_0$ when it becomes divergent with large enough opening. 

Let us assume that the flow variables obey: 
\begin{eqnarray}
|u_{1,y}| \ll |u_{1,x}|, \; |u_{2,y}| \ll |u_{1,x}|, \; {\rm away\; from \;
boundaries\; of\;} \Omega_0,  \nonumber \\
|\vec{u}_{y}| \gg |\vec{u}_{x}|, \; {\rm near\; the\; boundaries\; of\;}
\Omega_0,  \nonumber \\
|\rho_y| \ll |\rho_x|, \; {\rm throughout\;} \Omega_0.  \label{A7}
\end{eqnarray}
These are consistent with physical observations in the viscous boundary
layers (Batchelor \cite{Batch}, page 302), 
namely, there are large vertical velocity gradients,
yet small pressure or density gradients in the boundary layers. 
The boundary layers are of width $O(\mu^{1/2})$. 
Denote by $\overline{\rho}$, $\overline{u%
}_{1}$, the vertical averages of $\rho$ and $u_1$. Note that the exterior
normal $\vec{n} = (-A_x/2,1)/(1+A_x^2/4)^{1/2}$ if $y=A/2$, $\vec{n} =
(-A_x/2,-1)/(1+A_x^2/4)^{1/2}$ if $y=- A/2$.

\noindent Let $a=x$, $b=x+\mbox{$\delta$} x$, $\mbox{$\delta$} x \ll 1$, 
$t$ slightly larger than $t_0$. We have:
\begin{equation}
{\frac{d}{dt}}\int_{\Omega (t)} \, \rho\, dV = {\frac{d}{dt}}\int_{\Omega
(t_0)} \, \rho\, J(t)\, dV = \int_{\Omega (t_0)}\, \rho_t \, J(t)\, dV +
\int_{\Omega (t_0)}\, \rho\, J_t\, dV,  \label{J1}
\end{equation}
where $J(t)$ is the Jacobian of volume change from a reference time $t_0$ to 
$t$. Since $\Omega(t)$ is now a thin slice, $J(t) ={\frac{A(t) }{A(t_0)}}$
for small $\mbox{$\delta$} x$, and $J_t = A_{t}(t)/A(t_0)$. The second
integral in (\ref{J1}) is: 
\begin{equation}
\int_{\Omega (t_0)}\, \rho\, J_t\, dV = \overline{ \rho} 
{\frac{A_{t}(t)}{A(t_0)}}
A(t_0) \mbox{$\delta$} x = \overline{\rho} \, A_{t}(t) \mbox{$\delta$} x.
\label{J2}
\end{equation}
The first integral is simplified using (\ref{A5}) as: 
\begin{equation}
\int_{\Omega (t_0)}\, \rho_t \, J(t)\, dV = \int_{\Omega (t)} \, \rho_t\, dV
= - \int_{\partial \Omega (t)} \rho \, \vec{u}\cdot \vec{n}\, dS.  \label{J3}
\end{equation}
We calculate the last integral of (\ref{J3}) further as follows: 
\begin{eqnarray}
\int_{\partial \Omega} \rho \vec{u}\cdot \vec{n} \, ds 
& = & \int_{-A/2}^{A/2}\,
(-\rho u_{1})(x,y,t) \, dy + \int_{-A/2}^{A/2}\, (\rho u_{1})(x+%
\mbox{$\delta$} x,y,t) \, dy  \nonumber \\
& + & \int_{x}^{x+\mbox{$\delta$} x}\, \rho\cdot (0,A_t/2)\cdot
(-A_{x}/2,1)\, dx  \nonumber \\
&+ & \int_{x}^{x+\mbox{$\delta$} x}\, \rho\cdot (0,-A_t/2)\cdot
(-A_{x}/2,-1)\, dx  \nonumber \\
& = & \overline{\rho u_{1}} A|^{x+\mbox{$\delta$} x}_{x} + {\frac{%
\mbox{$\delta$} x}{2}}(\rho\, A_{t})|_{y=A/2} + {\frac{\mbox{$\delta$} x}{2}}%
(\rho\, A_{t})|_{y=-A/2}+ O((\mbox{$\delta$} x)^2)  \nonumber \\
& \approx & (\overline{\rho}\cdot \overline{u}_{1} A)|^{x+\mbox{$\delta$}
x}_{x}+ \overline{\rho}A_{t} \mbox{$\delta$} x + O((\mbox{$\delta$} x)^2),
\label{A9}
\end{eqnarray}
where we have used the smallness of $\rho_{y}$ to approximate $\rho|_{y=\pm
A/2}$ by $\overline{\rho}$ and $\overline{\rho u_{1}}$ by $\overline{\rho}%
\cdot \overline{u_{1}}$.

\noindent Combining (\ref{J1})-(\ref{J3}), (\ref{A9}) with: 
\begin{equation}
{\frac{d}{dt}} \int_{\Omega} \rho \, dV = (\overline{\rho} A \mbox{$\delta$}
x)_{t} + O((\mbox{$\delta$} x)^2),  \label{A8}
\end{equation}
dividing by $\mbox{$\delta$} x$ and sending it to zero, we have: 
\[
(\overline{\rho}A)_{t} + (\overline{\rho}\cdot \overline{u}_{1} A)_{x}  = 0, 
\]
which is (\ref{E1}).

Next consider $i=1$ in the momentum equation, $a=x$, $b=x+\mbox{$\delta$} x$%
. We have similarly with (\ref{E6}): 
\begin{eqnarray}
{\frac{d }{dt}} \int_{\Omega (t)} \, \rho\, u_1\, dV  =  \int_{\Omega (t)}
(\rho u_{1})_{t}\, dV + \int_{\Omega (t_0)} \rho\, u_1\, J_t\, dV  \nonumber
\\
 = - \int_{\partial \Omega (t)}\rho u_{1} \vec{u}\cdot \vec{n}\, dS +
\int_{\partial \Omega (t)} \sigma_{1,j}\cdot \vec{n}_{j}\, dS 
 + \overline{\rho}\, \overline{u_1}\, A_{t} %
\, \mbox{$\delta$} x + O(\delta x^2).  \label{J4}
\end{eqnarray}

\noindent We calculate the integrals of (\ref{J4}) below. 
\be
{\frac{d }{dt}}\int_{\Omega } \rho u_{1}\, dV  = 
 (\overline{\rho}\overline{u}
_{1}A)_{t}\mbox{$\delta$} x + O((\mbox{$\delta$} x)^2) \approx (\overline{%
\rho}\cdot \overline{u} A)_{t}\cdot \mbox{$\delta$} x + O((\mbox{$\delta$}
x)^{2}). \label{key1}
\ee
{\bf Using $u_1=0$ on the upper and lower boundaries}, a similar calculation 
as (\ref{A9}) gives:
\be
\int_{\partial \Omega}\rho u_{1} \vec{u}\cdot \vec{n}\, dS = (\overline{\rho}%
\cdot \overline{u}_{1}^{2}A)|_{x}^{x+\mbox{$\delta$} x} + 
O(\dta x \, \mu^{1/2}),
\label{conv}
\end{equation}
where the smallness of $u_{1,y}$ in the interior and small width of boundary
layer $O(\mu^{1/2})$ gives the $O(\mu^{1/2})$ for approximating $\overline{%
u_{1}^2}$ by $\overline{u_1}\cdot \overline{u_{1}}$.

\begin{rmk}
 Notice that for inviscid flows, 
we would have an additional term $\overline{\rho}\,
\overline{u}_1\,A_t\,\dta x$, which would cancel the third term on the 
right hand side of (\ref{J4}). As a result, the $A_t\, u/A$ term would 
be absent from the momentum equation (\ref{E2}).
\end{rmk}

Let us continue to calculate: 
\begin{eqnarray}
\int_{\partial \Omega} - p \mbox{$\delta$}_{1,j}n_{j} dS &\approx & -%
\overline{p}A|_{x}^{x+\mbox{$\delta$} x} + \int_{x}^{x+\mbox{$\delta$} x}
p\, A_{x}\, dx  \nonumber \\
& = & -\overline{p}A|_{x}^{x+\mbox{$\delta$} x} + \overline{p}\, A_{x}\, %
\mbox{$\delta$} x + O( (\mbox{$\delta$} x)^{2}).  \nonumber
\end{eqnarray}

\noindent Noticing that:
\[
d_{11} = 2\mu (u_{1,x} - (u_{1,x} + u_{2,y})/3 ),\; 
d_{12} = 2\mu (u_{1,y}+u_{2,x}). 
\]
It follows that 
\[ \overline{
d_{11}} = {\frac{4}{3}}\mu \overline{u}_{1,x} - {2\mu A_{t}\over 3 A}.\]
Thus the
contribution from the left and right boundaries located at $x$ and $x +%
\mbox{$\delta$} x$ is: 
\begin{equation}
\sum_{l,r}\int_{l,r} \, d_{11}n_{1} =  A\, \overline{d_{11}}|_{x}^{x+%
\mbox{$\delta$} x} = {\frac{4}{3}}A \, \mu\, \overline{u}_{1,x}|_{x}^{x+%
\mbox{$\delta$} x} - {2\mu \, A_{t}\over 3}|^{x+\delta x}_{x}.
\label{A11}
\end{equation}
The contribution from the upper and lower boundaries is: 
\begin{eqnarray}
\sum_{\pm}\int_{y=\pm A/2} d_{11}\, n_{1}\, dS \, & = & - d_{11}A_{x}%
\mbox{$\delta$} x/2 |_{y=A/2} - d_{11}A_{x}\mbox{$\delta$} x/2 |_{y=-A/2} 
\nonumber \\
& = & \mu \mbox{$\delta$} x \sum _{\pm} O(\partial_{y} \vec{u})|_{y=\pm
A/2}.  \label{A12}
\end{eqnarray}
Similarly, 
\begin{equation}
\sum_{\pm}\int_{y=\pm A/2}d_{12} n_{2}\, dS = \mu \mbox{$\delta$} x
\sum_{\pm} O(\partial_{y}\vec{u})|_{y=\pm A/2}.  \label{A13}
\end{equation}

Since $\partial_{y} \vec{u}|_{y=\pm A/2} =O(\mu^{-1/2})$, the viscous
flux from the boundary layers are $O(\dta x \mu^{1/2})$,  much larger than the
averaged viscous term $\dta \, x\, 
{\frac{4\mu }{3}}(A\overline{u_{1}}_{x})_{x}= O(\dta \, x \, \mu)$.
We notice that the
vertically averaged quantities have little dependence on the viscous
boundary layers unless $A$ is on the order $O(\mu^{1/2})$. 
Hence the quantities from upper and lower edges in (\ref
{A12}) and (\ref{A13}), and that in (\ref{conv}), should balance themselves.
Omitting them altogether, and combining remaining terms that involve only $%
\overline{u_1}$, $\overline{\rho}$ in the bulk, we end up with 
(after dividing by $\delta x$ and sending it to zero): 
\begin{equation}
(\overline{\rho}\cdot \overline{u_1} A)_{t} +(\overline{\rho}\cdot \overline{%
u_1}^2 A)_{x} = -(\overline{p}\, A)_{x} + A_{x}\overline{p} 
+\overline{\rho}\, \overline{u_1}\, A_{t} 
+ {\frac{4\mu}{3}}
(A\overline{u_1}_{x})_{x} - 2\mu A_{tx}/3,  \label{A14}
\end{equation}
which gives (\ref{E2}) in the inviscid limit $\mu \ra 0$.

\newpage

\begin{center}
Two Mass Model Parameters in cgs Unit.
\end{center}
\ba
m_1  & = &  0.17 \; g \no \\
m_2    & = & 0.03 \; g \no \\
x_2 -x_1 &=&    0.2\; cm \no\\
x_1 + L   & = &  0.025 \; cm \no \\
k_{1,open} &= &  45 \; kdynes \no \\
k_{1,closed} &=&   180 \; kdynes \no \\
y_{0,1} &=&   0 \; cm \no \\
k_{2,open} &=&  8 \; kdynes \no \\
k_{2,closed} &=&  32 \; kdynes \no \\
y_{0, 2}  &=&  0.0 \; cm \no \\
k_{12}  &=& 25 \; kdynes \no \\
y_{0,12} &=& 0 \; cm \no \\
y_c  &=& 0.001 \; cm \no \\
A(-L,t) &=&   2 \; cm \no \\
r_{1,open} &=&   17.5 \; dynes/(cm \; s) \no \\
r_{1,closed}  &=&  192.4 \; dynes/(cm \; s)  \no \\
r_{2,open} &=&  18.6 \; dynes/(cm \; s) \no \\
r_{2,closed} &=&   49.6 \; dynes/(cm \; s).
  \no \ea
 
\newpage
\bigskip

\medskip

\medskip

\medskip

\medskip

\medskip

\end{document}